\documentclass[journal,onecolumn,10pt,twoside]{IEEEtran}
\normalsize

\usepackage{graphicx,amssymb,amsthm}

\usepackage{tikz,url}
\usepackage[cmex10]{amsmath}
\newtheorem{theorem}{Theorem}
\newtheorem{remark}{Remark}
\newtheorem{lemma}{Lemma}

\newtheorem{assumption}{Assumption}
\newtheorem*{definition}{Definition}

\newtheorem{cor}{Corollary}

\usepackage{cite}
\usepackage{color}
\usepackage{graphicx}
\usepackage{array}

\usepackage[tight,footnotesize]{subfigure}
\usetikzlibrary{arrows,decorations.pathmorphing,backgrounds,fit,positioning,shapes.symbols,chains}

\DeclareMathAlphabet{\mathpzc}{OT1}{pzc}{m}{it}

\makeatletter

\DeclareMathOperator*{\argmax}{argmax}                                          

\begin{document}
 
\title{Optimal Communication Strategies in Networked Cyber-Physical Systems with Adversarial Elements}

\author{Emrah~Akyol,
        Kenneth~Rose,
        Tamer~Ba\c{s}ar, and C\'edric Langbort  
\thanks{E. Akyol, T. Ba\c{s}ar and C. Langbort are with the  Coordinated Science Laboratory, University of Illinois at Urbana-Champaign, 1308
       West Main Street, Urbana, IL 61801, USA email: \{akyol, basar1, langbort\}@illinois.edu.  K. Rose is with the Department
of Electrical and Computer Engineering, University of California, Santa Barbara,
CA, 93106 USA e-mail: rose @ece.ucsb.edu. }

\thanks{The material in this paper was presented
in part at the IEEE International Symposium on Information Theory (ISIT), Turkey, July 2013 and at the  Conference on Decision and Game Theory for Security(GameSec) Nov. 2013, Forth Worth, Texas, USA.}

\thanks{This work was supported in part by  NSF under grants CCF-1016861,  CCF-1118075, CCF-1111342, CCF-1320599 and also by an Office of Naval Research (ONR) MURI Grant N00014-16-1-2710.}
}


\maketitle


\begin{abstract}
This paper studies optimal communication and coordination strategies in cyber-physical systems for both defender and attacker within a game-theoretic framework.  We  model the communication network of a cyber-physical system as a sensor network which involves one single Gaussian source observed by many sensors, subject to additive independent Gaussian observation noises. The sensors communicate with the estimator over a  coherent Gaussian multiple access channel. The aim of the receiver is to reconstruct the underlying source with minimum mean squared error. The scenario of interest here is one where some of the sensors are captured by the attacker and they act as the adversary (jammer): they strive to maximize distortion. The receiver (estimator) knows the captured sensors but still cannot simply ignore them due to the multiple access channel, i.e., the outputs of all sensors are summed to generate the estimator input.  We show that the ability of transmitter sensors to secretly agree on a random event, that is ``coordination", plays a key role in the analysis. Depending on the coordination capability of the sensors and the receiver,  we consider three different problem settings.  The first setting involves transmitters and the receiver  with ``coordination" capabilities. Here, all transmitters can use identical realization of randomized encoding for each transmission.  In this case, the optimal strategy for the adversary sensors also exploits coordination, where they all generate the same realization of independent and identically distributed Gaussian noise.  In the second setting, the transmitter sensors are restricted to use  deterministic encoders, and this setting, which corresponds to a Stackelberg game, does not admit a saddle-point solution. We  show that  the optimal strategy for all sensors is uncoded communications where encoding functions of adversaries and transmitters are aligned in opposite directions. In  the third, and last, setting where only a subset of the transmitter and/or jammer sensors can coordinate, we show that  the solution radically depends on the fraction of the transmitter sensors that can coordinate. 
In the second half of the paper, we extend our analysis to an asymmetric scenario where we remove the assumption of identical power and noise variances for all sensors. Limiting the optimal strategies to conditionally affine mappings, we derive the optimal power scheduling over the sensors. We show that optimal power scheduling renders coordination superfluous for the attacker, when the transmitter sensors exploit coordination, as the attacker allocates all adversarial power to one sensor. In the setting where coordination is not allowed, both the attacker and the transmitter sensors distribute power among all available sensors to utilize the well-known estimation diversity in distributed settings. 
\end{abstract}


\section{Introduction}

Cyber-physical systems (CPSs) are large-scale interconnected systems of heterogeneous, yet collaborating,
components that  provide integration of computation with physical processes \cite{kim2012cyber}. The inherent heterogeneity and integration of different components in  CPS pose new security challenges\cite{sandberg2015cyberphysical}. One such security challenge pertains to the CPS communication network. 

Most CPSs rely on the presence of a Wireless Sensor Network (WSN)
composed of distributed nodes that communicate their measurements to a central state estimator (fusion center) with 
higher computation capabilities. Efficient and reliable communication of these measurements is a critical aspect of WSN
systems that determine usability of the infrastructure. Consider the architecture shown in Figure 1 where multiple sensors observe the state of the plant and transmit their observations over a wireless multiple access channel (MAC) to a central estimator (fusion center) which decides on the control action. The sensors in such architectures are known to be  vulnerable to various attacks, see e.g., \cite{fawzi2014secure,pasqualetti2013attack,mo2014detecting} and the references therein. For example, sensors may be captured and analyzed such that the attacker
gains insider information about the communication scheme and networking protocols. The attacker can then reprogram the
compromised sensors and use them to launch the so-called Byzantine attack\cite{lamport1982byzantine,dolev1982byzantine,6582732,kosut2008distributed}, where the objective of these adversarial sensors can be i) to distort the estimate made at the fusion center, which corresponds to a zero-sum game where the transmitting sensors aim to minimize some distortion associated the state measurements while the objective of the attacker is to maximize it, or  ii) strategically craft messages to deceive the estimator in a way to render its estimate close to a  predetermined, biased value\cite{akyol}, as was done in the replay attacks of StuxNet in SCADA systems \cite{langner2011stuxnet}. This paper presents an information/communication theoretic approach to  Bayesian optimal sensor fusion in the presence of Byzantine sensors for the first setting, while a preliminary analysis of the second case can be found in \cite{asilomar16}.

We analyze the communication scenario from the perspective of joint source-channel coding (JSCC) which has certain advantages over separate source and channel coding for sensor networks; see e.g.,  \cite{gastpar2005power} and the references therein. In this paper, we extend the game theoretic analysis of the Gaussian test channel \cite{basar1983gaussian,basar1985complete,basar1986solutions,bansal1989communication} to Gaussian sensor networks studied by  \cite{gastpar2005power,xiao2008linear,4915748,1597575,1657815,4568456,6731588,5089496}.  In \cite{xiao2008linear}, the performance of a simple uncoded communication is studied,  in conjunction with optimal power assignment over the sensors given a sum power budget. For a particular symmetric setting, Gastpar showed that indeed this uncoded scheme is optimal over all encoding/decoding methods that allow arbitrarily high delay \cite{gastpar2008uncoded}. However, it is well understood that in more realistic asymmetric settings, the uncoded communication scheme is suboptimal, and in fact, the optimal communication strategies are {\it unknown} for these settings\cite{ElGamalBook,lapidoth2010sending}.  
Information-theoretic analysis of the scaling behavior of such sensor networks, in terms of the number of sensors,  is provided  in \cite{leong2011scaling}.

In this paper, building on our earlier work on the topic \cite{akyol2013gaussian,akyol2013communication}, we consider three settings for the  sensor network model, which is illustrated in Figure 2 and described in detail in Section II. The first $M$ sensors (i.e., the transmitters) and the single receiver constitute Player 1 (minimizer) and the remaining $K$ sensors (i.e., the adversaries) constitute Player 2 (maximizer). Formulated as a zero-sum game, this setting does not admit a saddle point in pure strategies (deterministic encoding functions), but admits one in mixed  strategies (randomized functions).  In the first setting we consider, the transmitter sensors are allowed to use randomized encoders, i.e., all transmitters and the receiver agree on some (pseudo)random sequence, denoted as $\{\gamma\}$ in the paper. We coin the term ``coordination" for this capability, show that it plays a pivotal role in the analysis and the implementation of optimal strategies for both the transmitter and the adversarial sensors , and provide the mixed-strategy saddle-point solution in Theorem 1.    In the second setting, we have a hierarchical scheme; it can be viewed as a Stackelberg game where Player 1 is the leader, restricted to pure strategies, and Player 2 is the follower, who observes Player 1's choice of pure strategies and plays accordingly. We present  in Theorem 2 the optimal strategies for this Stackelberg game, whose cost is strictly higher than the cost associated with the first setting. The sharp contrast between the two settings underlines the importance of ``coordination" in sensor networks with adversarial nodes. In the third setting, we consider  only a given subset of the transmitters and also  the adversarial sensors can coordinate. We show that if the number of  transmitter sensors that can coordinate is sufficiently high (compared to ones that cannot), then the problem  becomes a zero-sum game with a saddle-point, where the coordination-capable transmitters use randomized linear strategy and the remaining transmitters are not used at all. It may at first appear to be counter intuitive to  forgo utilization of the second set of transmitter sensors but the gain from coordination (by the first set of transmitter sensors) more than compensates for this loss. Coordination is also important for the adversarial sensors. When transmitters coordinate, adversaries would benefit from coordination to generate identical realizations of Gaussian jamming noise. In contrast with transmitters, the adversarial sensors which cannot coordinate are of use: they generate independent copies of  identically distributed Gaussian jamming noise. Otherwise, i.e., the number of coordinating transmitters is not  sufficiently high, transmitters use deterministic (pure strategies) linear encoding, and optimal adversarial strategy is also uncoded communications  in the opposite direction of the transmitters.

\begin{figure}
\centering
\includegraphics[scale=0.3]{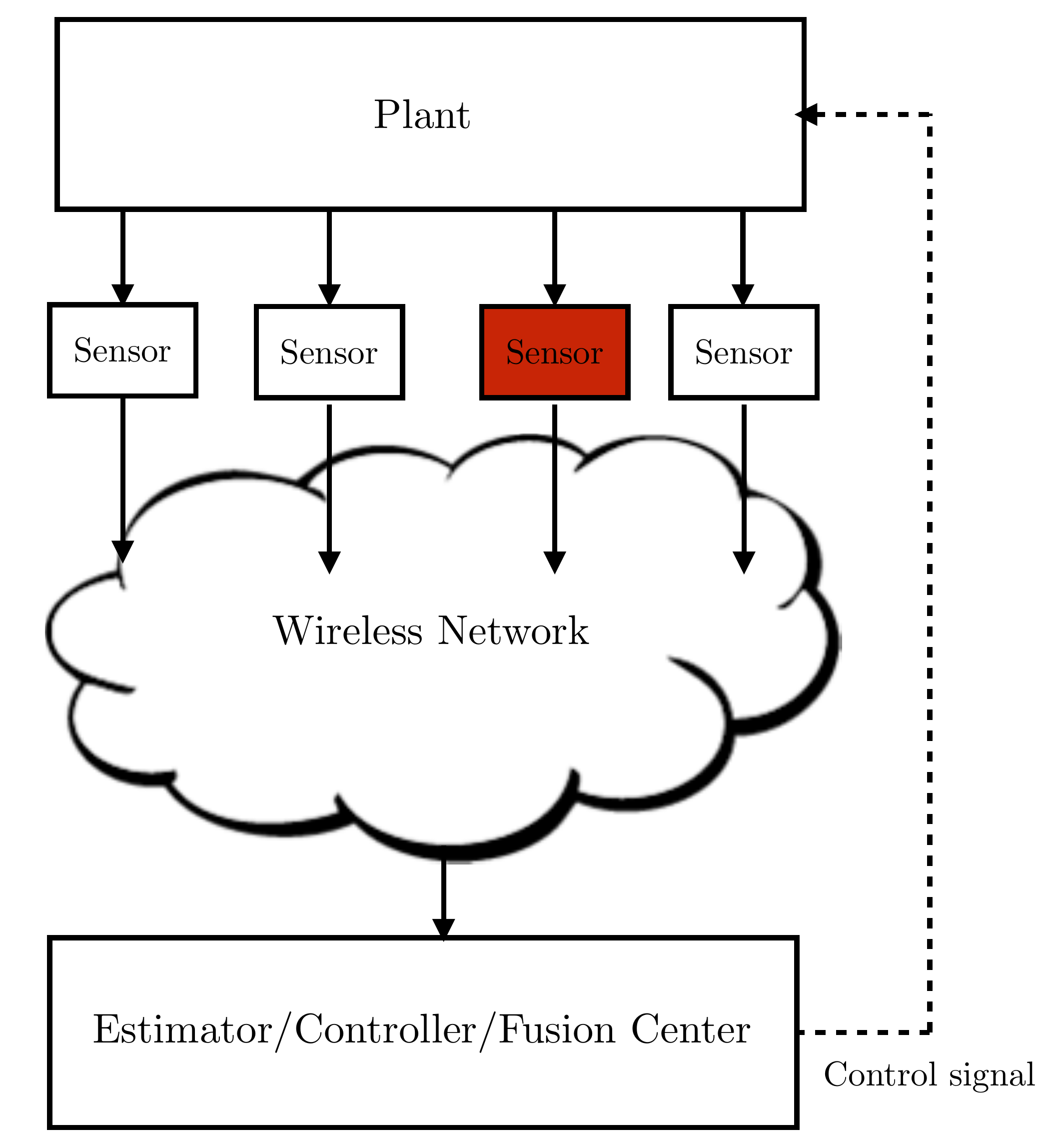}
\label{figure1}
\caption{The basic cyber-physical system model. The sensor in red color is a Byzantine sensor, i.e., it is  captured by the adversary.}
\end{figure}

In the second part of the paper, we extend the analysis to asymmetric settings where sensing and/or communications channels, and allowed transmission power of each sensor are different.  For  this setting,  information-theoretically optimal source-channel coding strategies are unknown (see e.g., \cite{akyol2016power} for inner and outer bounds of optimal performance). Here, we assume that the sensors use uncoded (zero-delay) linear communication strategies, which are optimal for the symmetric setting. We also allow another coordination capability to the sensors to combat with this inherent heterogeneity: we assume a total power limit over the sensors which allows for power allocation over sensors. We assume this power allocation optimization capability is also available  to the adversarial sensors. We derive optimal power scheduling strategies for the transmitter and the adversarial sensors for both settings, i.e., with or without coordination\footnote{Here, the term coordination refers to the sensors' ability on generating identical realization of a (pseudo)random sequence.}. We show that the power allocation capability renders coordination superfluous for the adversarial sensors, while it is still beneficial to the transmitter sensors.


This paper is organized as follows: In Section II, we formulate the problem. In Section III, we present our results pertaining to the symmetric setting, and in Section IV, we analyze the asymmetric case. In Section V, we present conclusions and discuss possible future directions of research. 

\begin{figure*}
\centering
\includegraphics[scale=0.55]{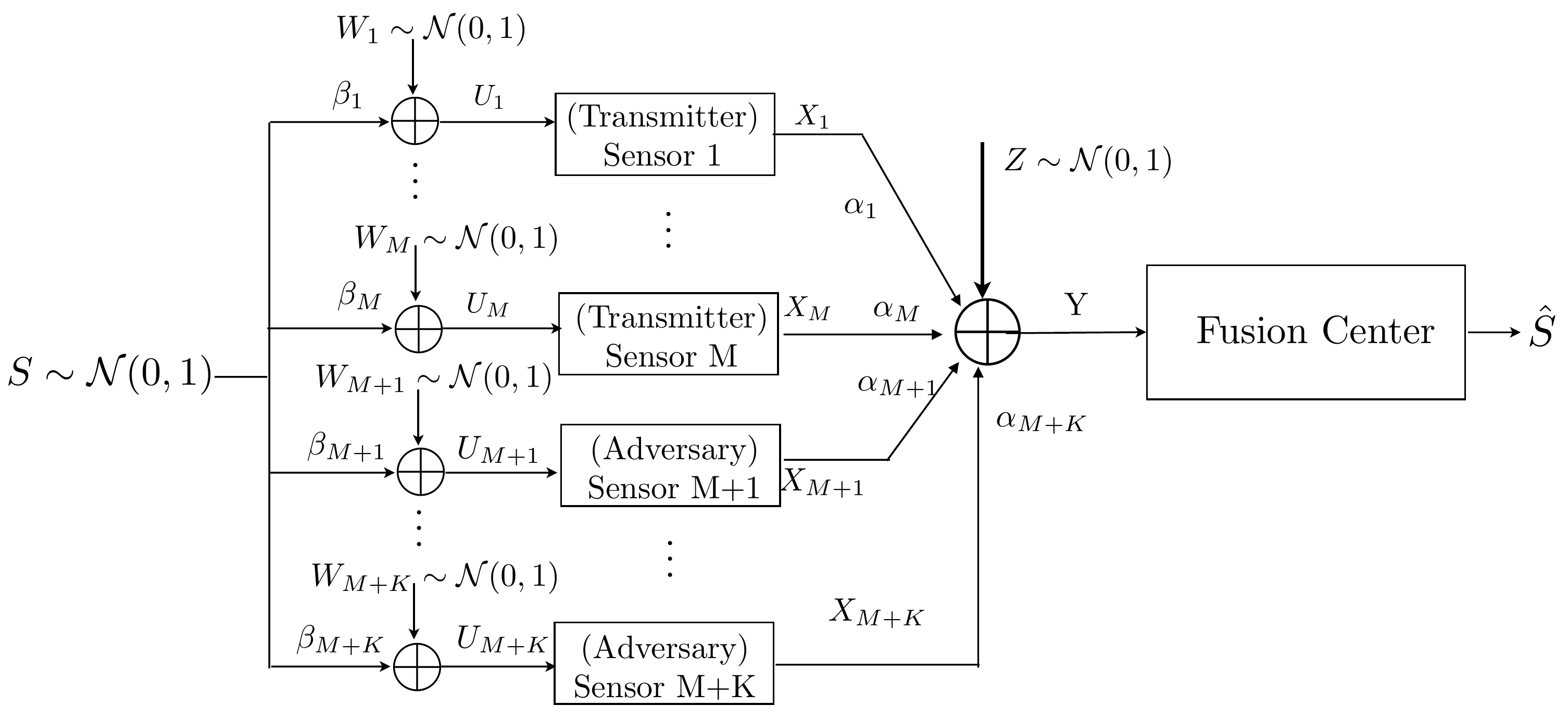}
\label{figure4}
\caption{The sensor network model.}
\end{figure*}

\section{Preliminaries}
\subsection{Notation}
 In general, lowercase letters (e.g., $x$) denote scalars, boldface lowercase (e.g., $\boldsymbol x$) vectors, uppercase (e.g., $U, X$) matrices and random variables, and boldface uppercase (e.g., $\boldsymbol X$) random
vectors. The $k^{th}$ element of vector $\boldsymbol x$ is denoted by $[\boldsymbol x]_k$. $\mathbb E(\cdot)$,  $\mathbb P(\cdot)$,  $\mathbb R$, and $\mathbb R^+$  denote, respectively,  the expectation and probability operators, and the sets of real and positive real numbers. $Bern(p)$ denotes the Binary random variable, taking values $1$ with probability $p$ and $-1$ with probability $1\!-\!p$. Gaussian distribution with mean vector $\boldsymbol \mu$ and covariance matrix $R$ is denoted as $\mathcal N(\boldsymbol \mu,R)$. The mutual information of random variables $X$ and $Y$ is denoted by $I(X;Y)$.

\subsection{Problem Formulation}
The sensor network model is depicted in Figure 2. The underlying source  $\{S(i)\}$ is a sequence of i.i.d. real valued Gaussian random variables with zero mean and unit variance\footnote{Normalizing the variance to 1 does not lead to any loss of generality.}. Sensor $m \in [1\!:\!M\!+\!K]$ observes a sequence $\{U_m(i)\}$  defined as 
\begin{equation}
U_m (i)= S(i)+\beta_m W_m(i),  
\end{equation}  
where $\{W_m(i)\}$ is a sequence of i.i.d.  Gaussian random variables with zero mean and unit variance, independent of  $\{S(i)\}$, and $\beta_m \in \mathbb R^+$ is the deterministic fading coefficient for the sensing channel. Sensor $m \in [1\!:\!M\!+\!K]$ can apply arbitrary Borel measurable function $g_m^N:\mathbb R^N \rightarrow \mathbb R^N$ to the observation sequence of length $N$, $\boldsymbol U_m$ so as to generate  the vector of length $N$  channel inputs $\boldsymbol X_m=g_m^N(\boldsymbol U_m)$ under power constraint: 
\begin{equation}
 \sum \limits_{i=1}^{N} \mathbb E \{X_m^2(i)\} \leq P_{m}
\label{power}
\end{equation}
The channel output is then given as 
\begin{equation}
Y(i)=Z(i)+\sum \limits_{m=1}^{M+K} \alpha_m\, X_m(i)
\end{equation}
where $\{Z(i)\}$ is a sequence of i.i.d. Gaussian random variables of zero mean and unit variance, independent of  $\{S(i)\}$ and  $\{W_m(i)\}$ and $\alpha_m \in \mathbb R^+$ is the deterministic fading coefficient for the communication channel of the $m$-th sensor. The receiver applies a Borel measurable function $h^N: \mathbb R^N \rightarrow \mathbb R^N$ to the received length-$N$ vector $\boldsymbol Y$ to generate $\boldsymbol {\hat S}$
\begin{equation}
\boldsymbol {\hat S} = h^N(\boldsymbol Y)
\label{costf}
\end{equation}
that minimize the cost, which is measured as mean squared error (MSE) between the underlying source $S$ and the estimate at the receiver $\hat S$ as
\begin{equation}
J(\{g_m^N(\cdot)\}_{m=1}^{M+K},h^{N}(\cdot))= \frac{1}{N}\sum \limits_{i=1}^{N} \mathbb E \{(S(i)-{\hat S}(i))^2\}.
\label{costf}
\end{equation}

{\bf Game model:} There are two players:  transmitter sensors and the receiver  constitute Player 1 who seeks to minimize  (\ref{costf})   over  $\{g_m^{N} (\cdot)\}_{m=1}^{M}$ and $h^N(\cdot)$.  Player 2 comprises the adversarial sensors whose common objective is to maximize  (\ref{costf})  by properly choosing  $\{g_k^N(\cdot)\}_{k=M+1}^{M+K}$. Since there is a complete conflict of interest, this problem constitutes a zero-sum game. We primarily consider the Stackelberg solution where Player-1 is the leader and plays first as a consequence of being the leader,  and the Player-2 is the follower, responds to the strategies of Player-1.  The game proceeds as follows: Player-1 plays first and announces its mappings. Player-2, knowing the mappings of Player-1, determines its own mappings that maximize t (\ref{costf}), given the strategy of Player 1. Player-1 of course, will anticipate this, and pick its mappings accordingly. The adversarial sensors have access to the knowledge of the strategy of the transmitter sensors (except the sequence of coordination variables $\{\gamma\}$ that enables Player-1 to use randomized strategies)  while the receiver has access to the strategies of all sensors, i.e., the receiver also knows the statistics of the sensors captured by the adversary.  We also note that the statistics of the variables, and the problem parameters, including the fading coefficients, are common knowledge.

More formally, we are primarily interested in 
\begin{equation}
J_U \triangleq \min \limits_{\{g_m^{N}\}_{m=1}^{M}, h^{N} } \max \limits_{\{g_k^{N}\}_{k=M+1}^{M+K}} J \left (\{g_m^{N}\}_{m=1}^{M},\{g_k^{N}\}_{k=M+1}^{M+K},h^{N} \right )
\end{equation}
which is the upper value of the game.

Some of the settings we analyze here admit, a special case of the described Stackelberg solution: a saddle-point solution. A transmitter-receiver-adversarial policy ($g_m^{N*},g_k^{N*},h^{N*}$) constitutes a saddle-point solution  if it satisfies the pair of inequalities
\begin{equation}
\label{saddle}
J(\{g_m^{N*}\}_{m=1}^{M},\{g_k^{N}\}_{k=M+1}^{M+K},h^{N*}) \leq J(\{g_m^{N*}\}_{m=1}^{M},\{g_k^{N*}\}_{k=M+1}^{M+K},h^{N*}) \leq J(\{g_m^{N}\}_{m=1}^{M},\{g_k^{N*}\}_{k=M+1}^{M+K},h^{N}) 
\end{equation} 

We also show that whenever a saddle-point solution exists, it is essentially unique\footnote{In these settings, multiple strategies, that are different only upto a sign change, yield the same cost. To account for  such trivially equivalent forms, we use the term ``essentially unique."}.  At the saddle point, it is well-known that the following holds (cf. \cite{basarbook}): 
 \begin{equation}J_U=J(\{g_m^{N*}\}_{m=1}^{M},\{g_k^{N*}\}_{k=M+1}^{M+K},h^{N*})=J_L\end{equation}
which we will refer as the saddle-point cost throughout the paper, where 
\begin{equation}
J_L=\max \limits_{\{g_k^{N}\}_{k=M+1}^{M+K}}  \min \limits_{\{g_m^{N}\}_{m=1}^{M}, h^{N} }  J \left (\{g_m^{N}\}_{m=1}^{M},\{g_k^{N}\}_{k=M+1}^{M+K},h^{N} \right )
\end{equation}
We are primarily concerned with the information-theoretic analysis of fundamental limits, and hence we take $N\rightarrow \infty$.

In this paper, we consider three different problem settings (denoted as settings I, II and III), depending on the ``coordination" capabilities of sensors. A salient encoding strategy that we will frequently encounter in this paper is the uncoded\footnote{Throughout this paper, we use ``uncoded", ``zero-delay"  interchangeably to denote ``symbol-by-symbol" coding structure. } linear communication strategy where the $N$-letter communication mapping $g_m^N$ consist of $N$ identical linear maps: $$g_m(U_m(i))=c_mU_m(i)$$ where $c_m$ satisfies the individual sensor power constraint with equality, i.e.,  $c_m=\sqrt{\frac{P_m}{1+\beta_m^2}}$ for $m=1, \ldots, M$.


%
%

\section{The Symmetric Scenario}
In this section, we focus on the symmetric scenario. More formally, we have the following symmetry assumption. 

\begin{assumption}[Symmetry Assumption] All sensors have identical problem parameters:   $\beta_m=\beta$,  $\alpha_m=\alpha$,  and $P_m=P$  for all $ m \in [1\!:\!M+K]$.
\end{assumption} 


\subsection{Problem Setting I}\label{sec1}

The first setting is concerned with the situation where the transmitter sensors have the ability to {\it coordinate}, i.e., all transmitters and the receiver can  agree on an i.i.d.  sequence of random variables $\{\gamma(i)\}$ generated, for example, by a side channel, the output of which is, however, not available to the adversarial sensors\footnote{An alternative practical method to coordinate is to generate the identical pseudo-random numbers  at each sensor, based on pre-determined seed.}. The ability of coordination allows transmitters and the receiver to agree on randomized encoding mappings. Perhaps surprisingly, in this setting, the adversarial sensors can also benefit from coordination, i.e., agree on an i.i.d. random sequence, denoted as $\{\theta(i)\}$, to generate the optimal jamming strategy.

The saddle-point solution of this problem is presented in the following theorem.

\begin{theorem}
Setting I, and under Assumption 1, admits a saddle-point solution with the following strategies: the strategy of the transmitter sensors is randomized uncoded transmission 
\begin{equation}
 X_m(i)=\gamma(i) \, c \, U_m(i), \,\,\, 1 \leq m\leq M
 \label{rand}
 \end{equation}
 where $\{\gamma(i)\}$ is an i.i.d. sequence of binary variables  $\gamma(i)\sim Bern (\frac{1}{2})$, and $c=\sqrt{\frac{P}{1+\beta^2}}$. 
The optimal jamming function (for adversarial sensors) is to generate the i.i.d. Gaussian output
 $$ X_k(i)=\theta(i), \,\,\,  M+1 \leq k\leq M+K $$
 where 
 $$\theta(i)\sim \mathcal N(0,  P),$$
and is independent of  the adversarial sensor input $U_k(i)$.
The strategy of the receiver is the Bayesian estimator of $S$ given $Y$, i.e.,
\begin{equation}
h(Y(i))=\frac{M \,c\, \alpha \beta}{ M^2 \alpha^2 \beta^2 c^2 +M c^2 \alpha^2+ {K^2P+1}}\gamma(i) \, Y(i).
\label{dec2}
 \end{equation}
The cost at this saddle-point is 
\begin{align}
J_{C}^S(M,K)=\frac{M c^2 \alpha^2+ {K^2P+1}}{ M^2 \alpha^2 \beta^2 c^2 +M c^2 \alpha^2+ {K^2P+1}}
\label{cost2}
\end{align}
\label{th2}
Moreover, this saddle-point solution is essentially unique. 
\end{theorem}
\begin{IEEEproof}
We start by verifying that the mappings given in the theorem satisfy the pair of saddle-point inequalities (\ref{saddle}), following the approach in \cite{basar1986solutions}.

RHS of (\ref{saddle}): Suppose the policy of the adversarial sensors is given as in Theorem \ref{th2}. Then, the communication system at hand becomes essentially identical to the problem considered in \cite{gastpar2008uncoded}, whose solution is uncoded communication with deterministic, linear encoders, i.e., $ X_m(i)= c U_m(i)$. Any probabilistic encoder, given in the form of (\ref{rand}) (irrespective of the density of $\gamma$) yield the same cost  (\ref{cost2}) with deterministic encoders and hence is optimal. Given that the optimal transmitter is in the form of (\ref{rand}), the optimal decoder is also zero-delay (symbol-by-symbol) mapping given as $h(Y(i))=\mathbb E\{S(i)|Y(i)\}=  \mathbb E \{SY\} (\mathbb E \{Y^2\})^{-1} Y(i)$ which can be explicitly obtained as in (\ref{dec2}) noting that   
\begin{align}
Y&=  c\, \alpha\,  \gamma(i) \sum \limits_{m=1} ^{M}  U_m(i) + \alpha \sum \limits_{k=M+1} ^{M+K}   \, X_k(i)+  Z(i) \\
\mathbb E \{SY\}&=\gamma(i)\, M  \beta  c  \alpha , \quad \mathbb E \{Y^2\}=1+K^2 P+ \left (M \beta  c  \alpha \right )^2+M \alpha^2 c^2 .
\end{align} 

The distortion is  (observing that $\sigma_S^2=1$):
 \begin{align}
J= 1- \frac{(\mathbb E \{SY\})^2}{\mathbb E \{Y^2\}}
= &\frac{M c^2 \alpha^2+ {K^2P+1}}{ M^2 \alpha^2 \beta^2 c^2 +M c^2 \alpha^2+ {K^2P+1}}
     \label{ascost}
 \end{align}

LHS of (\ref{saddle}): By the symmetry of the problem, we assume, without any loss of generality, that  all adversarial sensors use the same jamming strategy. Let us derive the overall cost conditioned on the realization of the transmitter mappings (i.e., $\gamma=1$ and $\gamma=-1$) used in conjunction with optimal linear decoders. If $\gamma=1$
\begin{equation}
D_1=J_1 +\xi \mathbb E\{SX_k\}+\psi \mathbb E\{ZX_k\}
\label{c1}
\end{equation}
  for some constants $\xi, \psi $, and similarly if  $\gamma=-1$
 \begin{equation}
D_2=J_1 -\xi \mathbb E\{SX_k\}-\psi  \mathbb E\{ZX_k\}
\label{c2}
\end{equation}
where the overall cost is
\begin{equation}
D(i)=\mathbb P(\gamma(i)=1)D_1+\mathbb P(\gamma(i)=-1)D_2 .
\end{equation} 
Clearly,  for $\gamma(i)\sim Bern (\frac{1}{2})$ the overall cost $J_1$  is only a function of the second-order statistics of the adversarial outputs, irrespective of the distribution of $\{\theta(i)\}$, and hence the solution presented here is indeed a saddle-point.
%
Having established this fact, we next show this saddle point is essentially unique.

Gaussianity of $X_k(i)$: The choice $X_k(i)=\theta(i)$ maximizes (\ref{costf}) since  it renders the simple uncoded linear mappings asymptotically optimal, i.e., the transmitters  cannot improve on the zero-delay performance by utilizing asymptotically high delays. Moreover, the optimal zero-delay performance is always lower bounded by the performance of the linear mappings, which is imposed by the adversarial choice of $X_k(i)=\theta(i)$. 

Independence of $\{X_k(i)\}$ of $\{S(i)\}$ and $\{W(i)\}$: If the adversarial sensors introduce some correlation, i.e., if $\mathbb E\{SX_k\}\neq0$ or $\mathbb E\{WX_k\}\neq0$, the transmitter can adjust its Bernoulli parameter to decrease the distortion. Hence, the optimal adversarial strategy is setting $\mathbb E\{SX_k\}=\mathbb E\{WX_k\}=0$ which implies independence since all variables are jointly Gaussian. 

Choice of Bernoulli parameter: Note that the optimal choice of the Bernoulli parameter for the transmitters is $\frac{1}{2}$ since other choices will not cancel the cross terms  in (\ref{c1}) and (\ref{c2}), i.e., $\mathbb E\{SX_k\}$ and $\mathbb E\{WX_k\}$. These cross terms can be exploited by the adversary to increase the cost, hence optimal strategy for transmitter is to set $\gamma=Bern(1/2)$.
\end{IEEEproof}

\begin{cor}[Value of Coordination]
Coordination, i.e., the ability of using a common randomized sequence, is beneficial to adversarial sensors in the case of coordinating transmitters and receiver, in the sense that  lack of adversarial coordination strictly decreases the overall cost. 
\end{cor}

\begin{IEEEproof}
Note that coordination, i.e., to be able to generate the same realization of $\theta(i)$ enables adversarial sensors to generate a Gaussian noise with variance $K^2P_A$  yielding the cost in (\ref{cost2}). However, without coordination, each sensor can only generate independent Gaussian random variables, yielding an overall Gaussian noise with variance $KP$ and the total cost 
\begin{equation}\frac{M c^2 \alpha^2+ {KP+1}}{ M^2 \alpha^2 \beta^2 c^2 +M c^2 \alpha^2+ {KP+1}}
 < J_{C}^S(M,K)
\end{equation} 
Hence, coordination of adversarial sensors strictly increases the overall cost.
\end{IEEEproof}


\begin{remark}
We note that the optimal strategies do not depend on the sensor index $m$, hence the implementation of the optimal strategy, for both transmitter and adversarial sensors, requires ``coordination" among the sensors. This highlights the need for coordination in game theoretic settings in sensor networks. Note that this coordination requirement arises purely from the game theoretic considerations, i.e., the presence of adversarial sensors. In the case where no adversarial node exists, transmitters do not need to ``coordinate". Moreover, as we will show in Theorem 2 if the transmitters cannot coordinate, then adversarial sensors do not need to coordinate. 
\end{remark}

%

\subsection{Problem Setting II}
Here, we address the second setting, where the transmitters do not have the ability to  secretly agree on a sequence of i.i.d. ``coordination" random variables, $\{\gamma\}$, to generate their transmission functions $X_m$. This setting does not admit a saddle-point solution, hence a Stackelberg solution is sought here. We also assume the number of adversarial sensors is less than the number of transmitter ones, i.e., $K<M$,  otherwise (if $M\geq K$) the adversarial sensors can effectively eliminate the output of the transmitters, and the problem becomes  trivial.

 We show that the essentially unique Stackelberg equilibrium is achieved by a transmitter strategy, which is  identical across all transmitters: uncoded transmission with linear mappings.  The equilibrium achieving strategy for the attack sensors, again identical across all adversarial sensors,  is uncoded transmission with linear mappings, but with the opposite sign of the transmitter. The receiver strategy is symbol-by-symbol optimal estimation of the source from the channel output. A rather surprising observation is that the adversarial coordination (in the sense of sharing a random sequence that is hidden from the transmitter sensors and  the receiver)  of  is superfluous for this setting, i.e., even if the adversarial sensors are allowed to cooperate, the optimal mappings and hence, the resulting cost at the saddle point do not change.

The following theorem captures this result.

\begin{theorem}
For setting II and under Assumption 1, the essentially unique Stackelberg equilibrium is achieved by: 
$$ X_m(i)=c\, U_m(i), \,\,\, 1 \leq m\leq M $$ 
for the transmitter sensors and
 $$ X_k(i)= -c \,U_k(i), \,\,\, M+1 \leq k\leq M+K $$
for the adversarial sensors.
The optimal receiver strategy is the symbol-by-symbol Bayesian estimator of $S$ given $Y$, i.e.,
\begin{equation}
h(Y(i))=\frac{(M-K) \,c\, \alpha \beta}{ (M-K)^2 \alpha^2 \beta^2 c^2 +(M-K) c^2 \alpha^2+ {1}}Y(i).
\label{dec4}
 \end{equation}
The cost at this Stackelberg solution is  
\begin{align}
J_{NC}^S(M,K)=\frac{(M-K) c^2 \alpha^2+ 1}{ (M-K)^2 \alpha^2 \beta^2 c^2 +(M-K) c^2 \alpha^2+ 1}
\label{cost4}
\end{align}

\end{theorem}
\begin{IEEEproof}
Let us first find the cost at the equilibrium, $J_{NC}^S$, for the given encoding strategies. We start by computing the expressions used in the  MMSE computations, 
\begin{align}
Y&=   c \, \alpha  \left (\sum \limits_{m=1}^{M}  U_m-\sum \limits_{k=M+1}^{M+K} U_k \right)  +  Z \\
\mathbb E \{SY\}&=(M-K) \,c\, \alpha \beta, \quad 
\mathbb E \{Y^2\}=(M-K)^2 \alpha^2 \beta^2 c^2 +(M-K) c^2 \alpha^2+ 1.
\end{align} 
Plugging these expressions, we obtain the cost, $J_{NC}^S(M,K)=1-\frac{(\mathbb E \{SY\})^2}{\mathbb E \{Y^2\}}$ as given in (\ref{cost4}), and the optimal receiver strategy, $h(Y(i))\mathbb E \{SY\} (\mathbb E \{Y^2\})^{-1} Y(i)$ as in (\ref{dec4}).  

We next show that linear mappings are the optimal (in the information-theoretic sense) encoding and decoding strategies. We first note that adversarial sensors have the knowledge of the transmitter encoding functions, and hence the adversarial encoding functions will be in the same form as the transmitters functions but with a negative sign i.e., since outputs are sent over an additive channel (see e.g., \cite{basar1986solutions,bansal1989communication} for a proof of this result). We next proceed to find the optimal encoding functions for the transmitters subject to this restriction. From the data processing theorem, we must have 
\begin{equation}
\label{eq}
I(\boldsymbol U_1, \boldsymbol U_2, \ldots, \boldsymbol U_{M+K}; \boldsymbol {\hat S}) \leq  I(\boldsymbol X_1, \boldsymbol X_2, \ldots, \boldsymbol X_{M+K}; \boldsymbol Y) 
\end{equation}
where we use the notational shorthand $\boldsymbol U_m=\left [ U_m (1), U_m(2),\ldots , U_m(N)\right]$  (and likewise for  $\boldsymbol X_m, \boldsymbol Y$ and $\boldsymbol {\hat S}$) for length $N$ sequences of random variables.  
 The left hand side can be lower bounded as:
\begin{align}
I(\boldsymbol U_1, \boldsymbol U_2, \ldots,\boldsymbol U_{M+K}; \boldsymbol {\hat S})\geq  R(D)
\end{align}
where $R(D)$ is the rate-distortion function of the Gaussian CEO problem adopted to our setting, and is derived in Appendix A. 
The right hand side can be upper bounded by 
\begin{IEEEeqnarray}{rCl}                                                                 
 I(\boldsymbol X_1, \boldsymbol X_2 ,..., \boldsymbol X_{M+K}; \boldsymbol Y)   
& \stackrel{\text{ (a)}}{\leq} & \sum \limits_{i=1}^{N}I(X_1(i), \ldots , X_{M+K}(i);Y(i)) \\
 &\leq  & \max \sum \limits_{i=1}^{N}I(X_1(i), \ldots , X_{M+K}(i);Y(i)) \label{aa}\\
 &=  & \frac{1}{2} \sum \limits_{i=1}^{N} \log ( 1+ { \boldsymbol 1}^T R_X(i) { \boldsymbol 1} )\label{bb}
 \end{IEEEeqnarray}
  where  $R_X(i)$ is  defined as 
\begin{equation}
\{R_X(i)\}_{p,r} \triangleq\mathbb E\{X_p(i)X_r(i)\} \quad \forall p,r \in [1:M\!+\!K].
\end{equation}
Note that (a) follows from the memoryless property of the channel and the maximum in (\ref{aa}) is over the joint density over $X_{1}(i), \ldots,X_{M+K}(i)$ given the structural constraints on $R_X(i)$ due to the power constraints. It is well known that the maximum is achieved, {\it uniquely}, by the jointly Gaussian density for a given fixed covariance structure \cite{diggavi2001worst}, yielding (\ref{bb}).
Since logarithm is a monotonically increasing function,  the optimal encoding functions $g_{m}^N(\cdot), m\in [1\!:\!M]$  equivalently maximize $\sum \limits_{p,r} \mathbb E\{X_p(i)X_r(i)\}$. Note that 
\begin{equation}
X_m(i)=\left [g_{m}^N(\boldsymbol U_m)\right ]_i
\end{equation}
and hence  $\{g_{m}^N(\cdot)\}_{m=1}^{M}$ that maximize 
\begin{equation}
\sum \limits_{p=1}^{p=M+K}\sum \limits_{r=1}^{r=M+K}  \mathbb E\{ [g_{p}^N(\boldsymbol U_p)]_i  [g_{r}^N(\boldsymbol U_r)]_i\}
\end{equation}
 can be found by invoking  Witsenhausen's lemma (given in Appendix B) as  $[g_{m}^N(\boldsymbol U_m)]_i=c\,\,   U_m(i)$ for all $i\in [1:N]$, and hence $g_{m}^N(\boldsymbol U_m)=c\,\,  \boldsymbol U_m$  for all $m\in [1:M]$. 
 Finally, we obtain $J_{NC}^S$ as an outer bound by equating the left and right hand sides of (\ref{eq}). The linear mappings in Theorem 2 achieve this outer bound, and hence are optimal. 


 \end{IEEEproof}

\begin{cor}
Source-channel separation, based on digital compression and communications is strictly suboptimal for this setting. 
\end{cor}
\begin{IEEEproof}
We first note that the optimal adversarial encoding functions must be the negative of that of the transmitters to achieve the saddle-point solution derived in Theorem 2. But then, the problem at hand becomes equivalent to a problem with no adversary which was studied in \cite{gastpar2005power}, where source-channel separation was shown to be strictly suboptimal. Hence, separate source-channel coding has to be suboptimal for our problem.  A more direct proof follows from the calculation of the separate source-channel coding performance.
\end{IEEEproof}

\begin{cor}
Coordination is beneficial to transmitter sensors, in the sense that  lack of  coordination strictly increases the equilibrium cost. 
\end{cor}
\begin{IEEEproof}
Proof follows from the fact that $J_C^S<J_{NC}^S$.
\end{IEEEproof}

\subsection{Problem Setting III}
The focus of this section is the setting between the two extreme scenarios of coordination, namely full, or no coordination. In the following, we assume that $M \epsilon$ transmitter sensors can coordinate with the receiver while $M(1-\epsilon)$ of them cannot coordinate, where $0<\epsilon<1$ and $M\epsilon$ is integer. Similarly, we consider only $K \eta$ of the adversarial sensors can coordinate while  the remaining $K (1-\eta)$ adversarial sensors cannot, where  $0<\eta<1$ and $K\eta$ is integer. Let us  reorder the sensors, without loss of generality, such that the first $M \epsilon$ transmitters and $K\eta$ adversaries can coordinate.  We again take $K<M$. Let us also define the quantity  $\epsilon_0$ as the unique\footnote{The fact that $J_C^S$ is monotonically decreasing in $\epsilon_0$ ensures that (\ref{pr3}) admits a unique solution.}  solution to:
\begin{equation}
J_C^S(M\epsilon_0, \sqrt{K^2\eta^2+K(1-\eta)})=J_{NC}^S(M, K)
\label{pr3}
\end{equation}

The following theorem captures our main result. 

\begin{theorem}
For $\epsilon > \epsilon_0 $, there exists a saddle-point solution  with the following strategies: the optimal transmission strategy requires that the  $M\epsilon$ capable transmitters use randomized linear encoding, while the  remaining $M (1-\epsilon)$ transmitters are not used. 
\begin{align}
 X_m(i)&=\gamma(i) \,c\, U_m(i), \,\,\, & 1 \leq m\leq M\epsilon \\
 X_m(i)&=0 \,\,\, &M\epsilon \leq m \leq M
 \label{rand555}
 \end{align}
where $\{\gamma(i)\}$ is an i.i.d. sequence of binary variables  $\gamma(i)\sim Bern (\frac{1}{2})$.  The optimal jamming policy (for the coordination-capable adversarial sensors) is to generate the identical Gaussian noise
 \begin{equation} X_k(i)=\theta(i), \,\,\,  M+1 \leq k\leq M+K \eta  \end{equation}
 while the remaining adversarial sensors will generate independent Gaussian noise 
 \begin{equation} X_k(i)=\theta_k(i), \,\,\,  M+K \eta \leq k\leq M+K\ \end{equation} where $\theta_k(i) \sim \mathcal N(0,  P)$
 are independent of  the adversarial sensor input $U_k(i)$.
The receiver strategy at this saddle point is 
\begin{equation}
h(Y(i))=\frac{M \epsilon\,c\, \alpha \beta}{ M^2 \epsilon^2 \alpha^2 \beta^2 c^2 +M\epsilon c^2 \alpha^2+ {(K^2 \eta^2+K(1-\eta)) P+1}}\gamma(i)\,Y(i).
\label{dec222}
 \end{equation}

If $\epsilon < \epsilon_0 $, the Stackelberg equilibrium is achieved with the deterministic linear encoding for the transmitter sensors, i.e., 
\begin{equation} X_m(i)=c\, U_m(i), \,\,\, 	1 \leq m\leq M\end{equation}
and the adversarial sensors use identical functional form with opposite sign of the transmitters, i.e.,
 \begin{equation} X_k(i)= -c\,U_k(i), \,\,\, M+1 \leq k\leq M+K \end{equation}
and the receiver uses 
\begin{equation}
h(Y(i))=\frac{(M-K) \,c\, \alpha \beta}{ (M-K)^2 \alpha^2 \beta^2 c^2 +(M-K) c^2 \alpha^2+ {1}}Y(i).
\label{dec455}
 \end{equation}
\end{theorem}

\begin{IEEEproof}
The transmitters have two choices: i) All transmitters will choose not to use randomization. Then, the adversarial sensors do not need to use randomization since the optimal strategy is deterministic, linear coding with the opposite sign, as shown in  Theorem 2.  Hence, the cost associated with this option is $J_{NC}^{S}(M,K)$. ii) Capable transmitters will use randomized encoding. This choice implies that remaining transmitters do not send information as they do not have access to randomization sequence $\{\gamma\}$, hence they are not used. The adversarial sensors which can coordinate generate  identical realization of the Gaussian noise while, remaining adversaries generate independent realizations. The total effective noise adversarial power will be $((K\eta)^2+(1-\eta)K)P$, and the cost associated with this setting is $J_C^S(M\epsilon, \sqrt{K^2\eta^2+K(1-\eta)})$. Hence,  transmitters will choose between two options depending on their costs, $J_C^S(M\epsilon, \sqrt{K^2\eta^2+K(1-\eta)})$ and $J_{NC}^S(M, K)$. Since, $J_C^S$ is a decreasing function in $M$ and hence in $\epsilon$, whenever $\epsilon >\epsilon_0$, transmitters use randomization (and hence so do the adversaries), otherwise problem setting becomes identical to ``no coordination". The rest of the proof simply follows from the proofs of Theorems 1 and 2. 
\end{IEEEproof}

\begin{remark}
Note that in the first regime ($\epsilon > \epsilon_0$),  we have a zero-sum game with saddle-point.  In the second regime ($\epsilon < \epsilon_0$), we have a Stackelberg game where all transmitters and receiver constitute the leader and adversaries constitute  the follower. 
\end{remark}

\begin{remark}
Theorem 3 states a rather interesting observation: depending on the network conditions, the optimal transmission strategy may  not use all of the transmitter sensors. At first glance, it might seem that discarding some of the available transmitter sensors is suboptimal. However, there is no feasible way to use these sensors, which cannot coordinate, without compromising the benefits of coordination.
\end{remark}

%

\section{The Asymmetric Scenario}
In this section, we remove the assumption  of identical sensing and channel noise variances and identical transmitter and adversary average power. Instead, we assume there is a sum-power limit for the set of transmitters and for the set of adversarial nodes.  In this general asymmetric case, the  optimal, in information-theoretic sense, communication strategies are unknown  in the absence of adversary. Here, we assume zero-delay linear strategies, in the light of our results in previous section, which provide an upper bound on the distortion-power performance: 
\begin{assumption}
In Setting-I (where the transmitter sensors can coordinate), the transmission strategies are restricted to 
\begin{equation} \label{mapping1}X_m(i)=\gamma (i)c_m U_m(i)\end{equation}
where $\{\gamma(i)\}$ is an i.i.d sequence of binary variables  $\gamma(i)\sim Bern (\frac{1}{2})$. 
In Setting-II (where no coordination is allowed),  the transmission strategies are limited to 
\begin{equation} \label{mapping2}X_m(i)=c_mU_m(i).\end{equation}
\end{assumption}
The problem we address in this section is two-fold: i) determine the optimal power allocation strategies for a given sum-power constraint of the form:
\begin{equation}    \sum \limits_{m=1}^M P_m \leq P_{T},\end{equation} and ii) determine the optimal adversarial sensor strategies subject to a sum power constraint: 
\begin{equation}    \sum \limits_{k=M+1}^{M+K}  P_k \leq P_{A},\end{equation} 
 Before deriving our results, we introduce  a few  variables. 
 \begin{definition}
We let $k^*$ be the index of an adversarial sensor having the best communication channel, i.e., 
 $$ k^*\triangleq\argmax_{k \in [M+1:M+K]} \alpha_k$$ and $P_A'$ is the associated received power $P_A'\triangleq \alpha_{k^*}^2 \,P_A\,$. In the case of multiple $k^*$s, we pick one arbitrarily. 
\end{definition}

\subsection{Setting-I}
In this setting, the transmitters can coordinate, similar to the setting studied in Section \ref{sec1}. Following the same steps as in Section  \ref{sec1}, we conclude that the solution sought here is a saddle point. 
\begin{theorem}
For setting I, and under Assumption 2, an essentially unique saddle-point solution exists with the following strategies: the  communication strategy for the transmitter sensor $m$ is given in (\ref{mapping1}) where 
$$c_m= \frac{\lambda_2 \alpha_m \beta_m }{ 2\left (1+\beta_m^2+\lambda_1 \alpha_m^2 \right )}, \quad  \lambda_1= \frac{P_T}{1+P_A'},$$
$$   \lambda_2= \sqrt{ \frac{4P_T}{\sum \limits_{m=1}^{M} \frac{(1+\beta_m^2) \alpha_m^2\beta_m^2}{(1+\beta_m^2+\lambda_1\alpha_m^2)^2}} }$$
The attacker uses only sensor $k^*$, and it generates i.i.d. Gaussian output
 $$ X_k^*(i)=\theta(i), \,\,\,  \text{where} \,\,\theta(i)\sim \mathcal N(0,  P_A),$$ 
which is independent of  the adversarial sensor input $U_{k^*}(i)$.
The  receiver is the Bayesian estimator of $S$ given $Y$, i.e.,
\begin{equation}
h(Y(i))=\frac{\left (\sum \limits_{m=1} ^{M} \beta_m c_m \alpha_m\right )\, \gamma(i) \,Y(i)}{1+P_A'+\left (\sum \limits_{m=1} ^{M} \beta_m c_m \alpha_m \right )^2+ \sum \limits_{m=1} ^{M}\alpha_m^2 c_m^2 }.
\label{dec6}
 \end{equation}
The cost at this saddle-point solution is 
\begin{align}
J_{C}^{AS}= \left (1+ \lambda_1 \sum \limits_{m=1}^{M}\frac{ \alpha_m^2 \beta_m^2 }{ 2\left (1+\beta_m^2+\lambda_1 \alpha_m^2 \right )} \right)^{-1}.
\label{cost6}
\end{align}

\label{th4}
\end{theorem}

\begin{IEEEproof}
 The existence of an essentially unique saddle-point solution follows from the same reasoning in Theorem 1. Let us take the transmission strategy as given in  the theorem statement and derive  the optimal attack strategy. Note that essentially, the attacker's role is limited to adding Gaussian noise subject to attack power $P_A$, the only remaining question here is how to allocate its power to the sensors. The objective of the attacker is to maximize the effective channel noise, i.e., to maximize:
$\sum \limits_{k=M} ^{M+K}\alpha_k \mathbb E \{\theta_k^2\}  $ subject to $\sum \limits_{k=M} ^{M+K}\mathbb E\{\theta_k^2\}\leq P_A$. The solution of this problem is simply: the attacker picks the best attack channel, i.e., the sensor with the largest $\alpha_k$, allocates all adversarial power on this sensor. 

Applying the optimal encoding map given in (\ref{mapping1}), and given adversary strategy we have the following auxillary expressions for the terms used in standard MMSE estimation.
\begin{align}
Y&= \gamma(i)\sum \limits_{m=1} ^{M}  c_m \, \alpha_m U_m(i) +\sum \limits_{k=M+1} ^{M+K}   \, \alpha_k X_k(i)+  Z(i) \\
\mathbb E \{SY\}&=\gamma(i)\sum \limits_{m=1} ^{M} \beta_m c_m \alpha_m, \quad \mathbb E \{Y^2\}=1+P_A'+ \left (\sum \limits_{m=1} ^{M} \beta_m c_m \alpha_m\right )^2+ \sum \limits_{m=1} ^{M}\alpha_m^2 c_m^2 .
\end{align} 

The distortion is  (observing that $\sigma_S^2=1$):
 \begin{align}
J=& 1- \frac{(\mathbb E \{SY\})^2}{\mathbb E \{Y^2\}}\\
= &1- \frac{\left (\sum \limits_{m=1} ^{M} \beta_m c_m \alpha_m\right )^2}{1+P_A'+\left (\sum \limits_{m=1} ^{M} \beta_m c_m \alpha_m \right )^2+ \sum \limits_{m=1} ^{M}\alpha_m^2 c_m^2 }, \\
     =& \frac{1+K^2 P+\sum \limits_{m=1} ^{M}\alpha_m^2 c_m^2}{1+P_A'+\sum \limits_{m=1} ^{M}\alpha_m^2 c_m^2+\left (\sum \limits_{m=1} ^{M} \beta_m c_m \alpha_m \right )^2}. 
     \label{ascost}
 \end{align}
Then, the problem is to determine $c_m$ that minimizes (\ref{ascost}) subject to the power constraint,  $\sum \limits_{m=1} ^{M} (1+\beta_m^2) c_m^2 \leq P_T$. We first note that  this problem is not convex in $c_m$. By changing the variables, we  convert this problem into a convex form which is analytically  solvable. First, instead of minimizing the distortion with a power constraint, we can equivalently minimize the power with a distortion constraint. Since distortion is a convex function of the total power (otherwise it can be converted to a convex problem by time sharing), there is no duality gap by this modification (cf. \cite{boyd2004convex}).   The modified problem is to minimize:
\begin{equation}
\sum \limits_{m=1} ^{M}(1+\beta_m^2) c_m^2,
\end{equation}
 subject to 
  \begin{equation}
\frac{1+P_A'+\sum \limits_{m=1} ^{M}\alpha_m^2 c_m^2}{1+P_A'+\sum \limits_{m=1} ^{M}\alpha_m^2 c_m^2+\left (\sum \limits_{m=1} ^{M} \beta_m c_m \alpha_m \right)^2} \leq J.
\end{equation}
  Note that 
\begin{equation}
\frac{1}{J}=1+\frac{\left (\sum \limits_{m=1} ^{M} \beta_m c_m \alpha_m \right)^2}{1+P_A'+ \sum \limits_{m=1} ^{M}\alpha_m^2 c_m ^2}
\end{equation}  
Next, we introduce a slack variable 
\begin{equation}
r=\sum \limits_{m=1} ^{M}\alpha_m \beta_m c_m.
\label{slack}
\end{equation} 
The optimization problem is to minimize
 \begin{equation}
 \sum \limits_{m=1} ^{M}(1+\beta_m^2) c_m^2,
\end{equation}
 subject to 
  \begin{equation}
1+P_A'+\sum \limits_{m=1} ^{M}\alpha_m^2 c_m^2 \leq(J^{-1}-1)^{-1} r^2,
\end{equation}
 and (\ref{slack}). This problem is convex in the variables $c_m$ and $r$. Hence, we construct the Lagrangian cost as 
  \begin{align}
J&=\sum \limits_{m=1} ^{M}\!\left (1+\beta_m^2\right ) c_m^2+\!\lambda_1 \!\left(\!1\!+P_A'+\!\sum \limits_{m=1} ^{M}\!\alpha_m^2 c_m^2\!-\!\frac{r^2}{(J^{-1}-1)} \right) +\lambda_2 \left(r-\sum \limits_{m=1} ^{M}\alpha_m \beta_m c_m\right),
\end{align} 
 where $\lambda_1 \in \mathbb R^+$ and $\lambda_2 \in \mathbb R$. The first-order conditions for stationarity of the Lagrangian yield:
  \begin{equation}
 \frac{\partial J}{\partial c_m}\!=\!2 c_m (1\!+\!\beta_m^2 )\!+\!2 \lambda_1 c_m \alpha_m^2 \!-\! \lambda_2 \alpha_m \beta_m\!=\!0,
\label{iki}
 \end{equation}
   \begin{equation}
 \frac{\partial J}{\partial r}=-2\lambda_1 (J^{-1}-1)^{-1} r + \lambda_2=0 ,
 \label{soniki}
 \end{equation}
 and we have (\ref{slack}) and 
   \begin{equation}
 1+P_A'+\sum \limits_{m=1} ^{M}\alpha_m^2 c_m^2 =\left (J^{-1}-1\right )^{-1} r^2.
 \label{son}
 \end{equation}
 From  (\ref{iki}), we have 
   \begin{equation}
c_m= \frac{\lambda_2 \alpha_m \beta_m }{ 2\left (1+\beta_m^2+\lambda_1 \alpha_m^2 \right )}.
\label{gg}
\end{equation}
Using (\ref{gg}) in (\ref{slack}), we have
   \begin{equation}
\frac{\lambda_2^2}{4\lambda_1}  \sum \limits_{m=1}^{M} \frac{\alpha_m^2\beta_m^2 }{\left (1+\beta_m^2+\lambda_1 \alpha_m^2 \right )}=1+P_A'+ \frac{\lambda_2^2}{4} \sum \limits_{m=1}^{M} \frac{\alpha_m^4\beta_m^2 }{\left (1+\beta_m^2+\lambda_1 \alpha_m^2 \right )^2}
\label{ggg}
\end{equation}
which simplifies to 
   \begin{align}
\lambda_1(1+P_A')&=\frac{\lambda_2^2}{4}  \sum \limits_{m=1}^{M} \frac{\alpha_m^2\beta_m^2 (1+\beta_m^2) }{\left (1+\beta_m^2+\lambda_1 \alpha_m^2 \right )^2} = \sum \limits_{m=1}^{M}  P_m= P_T \Rightarrow  \lambda_1= \frac{P_T}{1+P_A'}.
                \label{3222}
\end{align}
We also have $$P_T= \sum \limits_{m=1}^{M} (1+\beta_m^2)c_m^2=\frac{\lambda_2^2}{4}\sum \limits_{m=1}^{M} (1+\beta_m^2) \frac{\alpha_m^2\beta_m^2}{(1+\beta_m^2+\lambda_1\alpha_m^2)^2}\Rightarrow \lambda_2= \sqrt{ \frac{4P_T}{\sum \limits_{m=1}^{M} \frac{(1+\beta_m^2) \alpha_m^2\beta_m^2}{(1+\beta_m^2+\lambda_1\alpha_m^2)^2}} }$$
Plugging the expressions of $\lambda_1$ and $\lambda_2$ in (\ref{soniki}), we obtain the equilibrium cost. 
\end{IEEEproof}
 \begin{remark}
 If Assumption 1 is replaced with Assumption 2 in setting I, coordination becomes redundant for the attacker. This is because the optimal attack strategy uses only one sensor, and there is no need to coordinate (generate the same realization of $\theta(i)$). 
\end{remark}

\begin{remark}{\label{imp}}
The optimal strategies can be computed for each sensor in a decentralized manner. The central agent can compute the optimal values of $\lambda_1$ and $\lambda_2$ and then broadcast this information to all sensors. Next, each transmitter sensor can compute its own  mapping based on local parameters $\alpha_m$ and $\beta_m$ and the broadcasted global parameters $\lambda_1$ and $\lambda_2$.
\end{remark}

Finally, we analyze the asymmetric setting where the sensors are not allowed to coordinate. We characterize the policies achieving the Stackelberg equilibrium and associated cost in the following theorem.

\begin{theorem}
For setting II, and under Assumption 2, the  encoding functions for the transmitter and the adversarial sensors at the Stackelberg equilibrium are:
$$ X_m(i)=c_m\, U_m(i), \,\,\, 1 \leq m\leq M, \quad  X_k(i)= c_k \,U_k(i), \,\,\, M+1 \leq k\leq M+K $$
where 
\begin{align}
c_m=\frac{\lambda_4 \alpha_m \beta_m }{ 2\left (1+\beta_m^2+\lambda_3 \alpha_m^2 \right )},\quad c_k= \frac{\lambda_2 \alpha_k \beta_k }{ 2\left (1+\beta_k^2-\lambda_1 \alpha_k^2 \right )}.
\end{align}
and $\lambda_1 \in \mathbb R, \lambda_2 \in \mathbb R^+, \lambda_3\in \mathbb R, \lambda_4 \in \mathbb R^+$ are constants that satisfy the following equations:  
 $$\lambda_2=-\frac{2P_A+2\lambda_1\left (1-\sum \limits_{m=1} ^{M}\alpha_m^2 c_m^2 \right)}{ \sum \limits_{m=1} ^{M}\alpha_m \beta_m c_m},   \quad \left(\frac{P_A+\lambda_1\left (1-\sum \limits_{m=1} ^{M}\alpha_m^2 c_m^2 \right)}{ \sum \limits_{m=1} ^{M}\alpha_m \beta_m c_m} \right)^2 \sum \limits_{k=M+1} ^{M+K}\frac{ (1+\beta_k^2)\alpha_k^2 \beta_k^2 }{ \left (1+\beta_k^2-\lambda_1 \alpha_k^2 \right )^2}  = P_A.$$
 and 
 $${\lambda_4}{\lambda_1}=-{\lambda_2}{\lambda_3}, \quad 1=\frac{P_T}{\lambda_1} +\frac{P_A}{\lambda_3}, \quad  \frac{\lambda_4^2}{4}   \sum \limits_{m=1}^{M}\frac{ \alpha_m^2 \beta_m^2 (1+\beta_m^2) }{\left (1+\beta_m^2+\lambda_3 \alpha_m^2 \right )^2} =P_T.
$$
The optimal receiver is the Bayesian estimator of $S$ given $Y$, i.e.,
\begin{equation}
h(Y(i))=\frac{\sum \limits_{m=1} ^{M+K} \beta_m c_m \alpha_m}{1+\sum \limits_{m=1} ^{M+K}\alpha_m^2 c_m^2+\left (\sum \limits_{m=1} ^{M+K} \beta_m c_m \alpha_m \right)^2}\,
Y(i).
\label{dec455}
 \end{equation}
The cost at this Stackelberg equilibrium is 
\begin{align}
J_{NC}^{AS}(M,K)=\left (1+ \lambda_3 \sum \limits_{m=1}^{M}\frac{ \alpha_m^2 \beta_m^2 }{ 2\left (1+\beta_m^2+\lambda_3 \alpha_m^2 \right )}- \lambda_1 \sum \limits_{k=M+1}^{M+K}\frac{ \alpha_k^2 \beta_k^2 }{ 2\left (1+\beta_k^2-\lambda_1\alpha_k^2 \right )} \right)^{-1}.
\label{cost444}
\end{align}

\end{theorem}

\begin{IEEEproof}
Since Player 1 (the transmitter sensors and the receiver) is the leader of this Stackelberg game and the adversarial sensors are the followers, we first compute the best response of the attacker to the given transmitter strategy and associated receiver policy. By the reasoning in Theorem 2, we conclude that the best adversary strategy is to use linear maps as given in the theorem statement.
 In the following, we compute the optimal adversary coefficients, $c_k, k \in [M+1:M+K]$ as a function of $c_m, m\in [1:M]$.  We first compute the expressions used in the  MMSE computations as:
\begin{align}
Y&= \sum \limits_{m=1} ^{M+K}  c_m \, \alpha_m U_m +  Z \\
\mathbb E \{SY\}&=\sum \limits_{m=1} ^{M+K} \beta_m c_m \alpha_m, \quad 
\mathbb E \{Y^2\}=1+ \left (\sum \limits_{m=1} ^{M+K} \beta_m c_m \alpha_m\right )^2+ \sum \limits_{m=1} ^{M+K}\alpha_m^2 c_m^2 .
\end{align} 
 The objective of the attacker is to maximize 
\begin{equation}
J=1- \frac{(\mathbb E \{SY\})^2}{\mathbb E \{Y^2\}}=\frac{1+\sum \limits_{m=1} ^{M+K}\alpha_m^2 c_m^2}{1+\sum \limits_{m=1} ^{M+K}\alpha_m^2 c_m^2+\left (\sum \limits_{m=1} ^{M+K} \beta_m c_m \alpha_m \right)^2}.
\label{cost5}
\end{equation}
over $c_k, k \in [M+1:M+K]$ that satisfy
\begin{equation}
\sum \limits_{k=M+1} ^{M+K}(1+\beta_k^2) c_k^2 \leq P_A. 
\label{power_constraint}
\end{equation}
This problem is again non-convex, hence we follow the approach we used in the proof of Theorem 4: we first introduce a slack variable. 
\begin{equation}
r_K=\sum \limits_{k=M+1} ^{M+K}\alpha_k \beta_k c_k,
\label{slack5}
\end{equation} 
and apply the  KKT optimality conditions. The stationarity conditions applied to the following Lagrangian cost   \begin{align}
J_A&=\sum \limits_{k=M+1} ^{M+K}\!\left (1+\beta_k^2\right ) c_k^2+\!\lambda_1 \!\left(\!{(r_K+\sum \limits_{m=1} ^{M}\alpha_m \beta_m c_m)^2}{(J^{-1}-1)^{-1}}-\!1-\!\sum \limits_{m=1} ^{M+K}\!\alpha_m^2 c_m^2\! \right) +\lambda_2 \left(r_K-\sum \limits_{k=M+1} ^{M+K}\alpha_k \beta_k c_k\right),
\end{align} 
 where $\lambda_1 \in \mathbb R^+$ and $\lambda_2 \in \mathbb R$, yield
  \begin{equation}
 \frac{\partial J_A}{\partial c_k}\!=\!2 c_k (1\!+\!\beta_k^2 )\!-\!2 \lambda_1 c_k \alpha_k^2 \!-\! \lambda_2 \alpha_k \beta_k\!=\!0,
\label{ikiiki}
 \end{equation}
   \begin{equation}
 \frac{\partial J_A}{\partial r_K}=2\lambda_1 (J^{-1}-1)^{-1} (r_K+\sum \limits_{m=1} ^{M}\alpha_m \beta_m c_m) + \lambda_2=0 ,
 \label{sonikig}
 \end{equation}
 and we have (\ref{slack5}) and 
   \begin{equation}
 1+\sum \limits_{m=1} ^{M}\alpha_m^2 c_m^2 + \sum \limits_{k=M+1} ^{M+K}\alpha_k^2 c_k^2  =\left (J^{-1}-1\right )^{-1} (r_K+\sum \limits_{m=1} ^{M}\alpha_m \beta_m c_m)^2.
 \label{son}
 \end{equation}
From  (\ref{ikiiki}), we have 
   \begin{equation}
c_k= \frac{\lambda_2 \alpha_k \beta_k }{ 2\left (1+\beta_k^2-\lambda_1 \alpha_k^2 \right )}.
\label{gggg}
\end{equation}
Using (\ref{gggg}) in (\ref{slack5}), we have
   \begin{equation}
\frac{-\lambda_2}{2\lambda_1}\sum \limits_{m=1} ^{M}\alpha_m \beta_m c_m-\frac{\lambda_2^2}{4\lambda_1}  \sum \limits_{k=M+1}^{M+K} \frac{\alpha_k^2\beta_k^2 }{\left (1+\beta_k^2-\lambda_1 \alpha_k^2 \right )}=1+\sum \limits_{m=1} ^{M}\alpha_m^2 c_m^2 + \frac{\lambda_2^2}{4} \sum \limits_{k=M+1}^{M+K} \frac{\alpha_k^4\beta_k^2 }{\left (1+\beta_k^2-\lambda_1 \alpha_k^2 \right )^2}
\label{ggg}
\end{equation}
which simplifies to 
   \begin{align}
\frac{-\lambda_2}{2\lambda_1}\sum \limits_{m=1} ^{M}\alpha_m \beta_m c_m-1-\sum \limits_{m=1} ^{M}\alpha_m^2 c_m^2= \frac{\lambda_2^2}{4 \lambda_1 }  \sum \limits_{k=M+1}^{M+K} \frac{\alpha_k^2\beta_k^2 (1+\beta_k^2) }{\left (1+\beta_k^2-\lambda_1 \alpha_k^2 \right )^2} =P_A/\lambda_1
                \label{orr}
\end{align}
or 
   \begin{align}
-\lambda_2 \sum \limits_{m=1} ^{M}\alpha_m \beta_m c_m-2\lambda_1(1-\sum \limits_{m=1} ^{M}\alpha_m^2 c_m^2) =2P_A \Rightarrow \lambda_2=-\frac{2P_A+2\lambda_1\left (1-\sum \limits_{m=1} ^{M}\alpha_m^2 c_m^2 \right)}{ \sum \limits_{m=1} ^{M}\alpha_m \beta_m c_m} 
                \label{orr4}
\end{align}

Plugging (\ref{gggg}) in (\ref{power_constraint}), we have 
 \begin{align} \frac{\lambda_2^2}{4}  \sum \limits_{k=M+1} ^{M+K} \frac{ (1+\beta_k^2)\alpha_k^2 \beta_k^2 }{\left (1+\beta_k^2-\lambda_1 \alpha_k^2 \right )^2}=\left(\frac{P_A+\lambda_1\left (1-\sum \limits_{m=1} ^{M}\alpha_m^2 c_m^2 \right)}{ \sum \limits_{m=1} ^{M}\alpha_m \beta_m c_m} \right)^2 \sum \limits_{k=M+1} ^{M+K}\frac{ (1+\beta_k^2)\alpha_k^2 \beta_k^2 }{ \left (1+\beta_k^2-\lambda_1 \alpha_k^2 \right )^2}  = P_A
 \label{ggggg}
\end{align}
The unique positive solution of (\ref{ggggg}) provides the value of $\lambda_1$ and by (\ref{orr4}), $\lambda_2$ can be computed, once $\lambda_1$ is obtained. Having obtained the optimal $c_k, k  \in [M+1:M+K]$ values as a function of $P_A$ and $c_m, m \in [1:M]$, we next derive $c_m$ that minimize (\ref{cost5}) subject to
\begin{equation}
\sum \limits_{m=1} ^{M}(1+\beta_m^2) c_m^2 \leq P_T. 
\label{power_constraint2}
\end{equation}
Again, we modify the problem as to minimize $\sum \limits_{m=1} ^{M}(1+\beta_m^2) c_m^2$ subject to 
  \begin{equation}
1+\sum \limits_{m=1} ^{M+K}\alpha_m^2 c_m^2 \leq(J^{-1}-1)^{-1} \left (r+\sum \limits_{k=M+1} ^{M+K}\alpha_k \beta_k c_k \right)^2,
\end{equation}
and \begin{equation}
r=\sum \limits_{m=1} ^{M}\alpha_m \beta_m c_m.
\label{slack3}
\end{equation}
The stationarity conditions applied to the following Lagrangian cost   \begin{align}
J_T&=\sum \limits_{m=1} ^{M}\!\left (1+\beta_m^2\right ) c_m^2+\!\lambda_3 \!\left(1+ \!\sum \limits_{m=1} ^{M+K}\!\alpha_m^2 c_m^2- \left (r+\sum \limits_{k=M+1} ^{M+K}\alpha_k \beta_k c_k \right)^2{(J^{-1}-1)^{-1}} \right) +\lambda_4 \left(r-\sum \limits_{m=1} ^{M}\alpha_m \beta_m c_m\right),
\end{align} 
 for $\lambda_3 \in \mathbb R^+$ and $\lambda_4 \in \mathbb R$ yield
  \begin{equation}
 \frac{\partial J_T}{\partial r}=-2\lambda_3 (J^{-1}-1)^{-1} \left (r+\sum \limits_{k=M+1} ^{M+K}\alpha_k \beta_k c_k \right) + \lambda_4=0 ,
 \label{sonikigg}
 \end{equation}
 
  \begin{align}
 \frac{\partial J_T}{\partial c_m}\!&=\!2 c_m (1\!+\!\beta_m^2 )\!+\!2 \lambda_3 c_m \alpha_m^2 \!+2\lambda_3\!\sum \limits_{k=M+1} ^{M+K}\!\alpha_k^2 c_kc_k' -2\lambda_3 (J^{-1}-1)^{-1} \left (r+\sum \limits_{k=M+1} ^{M+K}\alpha_k \beta_k c_k \right)\sum \limits_{k=M+1} ^{M+K}\alpha_k \beta_k c_k' \!- \lambda_4 \alpha_m \beta_m\!\nonumber\\
 &=2 c_m (1\!+\!\beta_m^2 )\!+\!2 \lambda_3 c_m \alpha_m^2 \!+2\lambda_3\!\sum \limits_{k=M+1} ^{M+K}\!\alpha_k^2 c_kc_k' -\lambda_4\sum \limits_{k=M+1} ^{M+K}\alpha_k \beta_k c_k' \!- \lambda_4 \alpha_m \beta_m\!=\!0
\label{ikigg}
 \end{align}
 where $c_k'=\frac{\partial c_k}{\partial c_m}$ and  (\ref{ikigg}) follows from (\ref{sonikigg}). We also have (\ref{slack3}) and    \begin{align}
 1+\sum \limits_{m=1} ^{M}\alpha_m^2 c_m^2 + \sum \limits_{k=M+1} ^{M+K}\alpha_k^2 c_k^2  &=\left (J^{-1}-1\right )^{-1} \left (r+\sum \limits_{k=M+1} ^{M+K}\alpha_k \beta_k c_k \right)^2 \\
 &= \frac{\lambda_4}{2 \lambda_3}\left (r+\sum \limits_{k=M+1} ^{M+K}\alpha_k \beta_k c_k\right)
 \label{sonnn}
 \end{align}
 as necessary conditions of optimality. Comparing (\ref{sonnn}) and (\ref{sonikig}), we have 
\begin{equation}
\frac{\lambda_4}{\lambda_3}=-\frac{\lambda_2}{\lambda_1}. 
\label{soneqqq}
\end{equation}
 We next use (\ref{soneqqq}) to rewrite the terms involving $c_k'$: 
 \begin{equation}\label{latest}
 2\lambda_3\!\sum \limits_{k=M+1} ^{M+K}\!\alpha_k^2 c_kc_k' -\lambda_4\sum \limits_{k=M+1} ^{M+K}\alpha_k \beta_k c_k' =  \frac{\partial}{\partial c_m} \left(2\lambda_1\!\sum \limits_{k=M+1} ^{M+K}\!\alpha_k^2 c_k^2 +{\lambda_2}{} \sum \limits_{k=M+1} ^{M+K}\alpha_k \beta_k c_k\right )= \frac{\partial}{\partial c_m} P_A=0
\end{equation}
 Using (\ref{latest}) in (\ref{ikigg}), we obtain 
   \begin{equation}
c_m= \frac{\lambda_4 \alpha_m \beta_m }{ 2\left (1+\beta_m^2+\lambda_3 \alpha_m^2 \right )} 
\label{gggggg}
\end{equation}
Plugging (\ref{gggggg})  and (\ref{gggg}) in (\ref{sonnn}) and using (\ref{soneqqq}), we have 
   \begin{equation}
 1+ \frac{\lambda_4^2}{4}\sum \limits_{m=1} ^{M} \frac{\alpha_m^4 \beta_m^2 }{ \left (1+\beta_m^2+\lambda_3 \alpha_m^2 \right )^2}  + \frac{\lambda_2^2}{4}\sum \limits_{k=M+1} ^{M+K} \frac{ \alpha_k^4 \beta_k^2 }{ \left (1+\beta_k^2-\lambda_1 \alpha_k^2 \right )^2} 
 =\frac{\lambda_4^2}{4 \lambda_3}\sum \limits_{m=1} ^{M} \frac{\alpha_m^2 \beta_m^2 }{ \left (1+\beta_m^2+\lambda_3 \alpha_m^2 \right )}-\frac{\lambda_2^2}{4 \lambda_1}  \sum \limits_{k=M+1} ^{M+K} \frac{\alpha_k^2 \beta_k^2 }{ \left (1+\beta_k^2-\lambda_1 \alpha_k^2 \right )} \nonumber
\end{equation}
which yields, after algebraic manipulations, 
   \begin{equation}
 1=\frac{P_T}{\lambda_1} +\frac{P_A}{\lambda_3}
\label{derivation}
\end{equation}
We also have
 \begin{equation}
 \frac{\lambda_4^2}{4}   \sum \limits_{m=1}^{M}\frac{ \alpha_m^2 \beta_m^2 (1+\beta_m^2) }{\left (1+\beta_m^2+\lambda_3 \alpha_m^2 \right )^2} =P_T.
 \label{derivation2}
 \end{equation}
 The set of equations (\ref{orr4}, \ref{ggggg}, \ref{soneqqq}, \ref{derivation}, \ref{derivation2}) (essentially) uniquely characterizes the variables $\lambda_1, \lambda_2, \lambda_3$ and $\lambda_4$. Plugging these variables into (\ref{sonikigg}), we obtain the equilibrium cost. 
\end{IEEEproof}

\begin{remark}
We again observe that, as noted in Remark \ref{imp}, the optimal power allocation admits a decentralized implementation: a central agent can compute and broadcast the values of constants $\lambda_i, i=1, \ldots, 4$ and the sensors can implement optimal communication strategies using the local information $\alpha_m$ and $\beta_m$ and these universal constants. The same interpretation also holds for the Byzantine sensors. 
\end{remark}

%
%

\section{Discussion and Conclusion}
In this paper, we have conducted a game-theoretical analysis of joint source-channel communication over a Gaussian sensor network with Byzantine sensors.  Depending on the {\it coordination} capabilities of the sensors, we have analyzed three problem settings. The first setting allows coordination among the transmitter sensors, first for the totally symmetric case. Coordination capability enables the transmitters to use randomized encoders. The saddle-point solution to this problem is randomized uncoded transmission for  the transmitters and the coordinated generation of i.i.d. Gaussian noise for the adversarial sensors. In the second setting, transmitter sensors cannot coordinate, and hence they use fixed, deterministic mappings. The solution to this problem is shown to be uncoded communication with linear mappings for both the transmitter and the adversarial sensors, but with opposite signs. We note that coordination aspect of the problem is entirely due to game-theoretic considerations, i.e., if no adversarial sensors exist, the transmitters  do not need coordination. In the third setting, where only a fraction of sensors can coordinate, the solution depends on the number of transmitter and adversarial sensors that can coordinate. If the gain from coordination for the transmitter sensors and the receiver, is sufficiently high, only the coordination-capable transmitter sensors are used. Then, the problem simplifies to an instance of setting I, i.e, there exists a unique saddle-point solution achieved by randomized linear mappings as the transmitter and the receiver strategy and independent noise as the adversarial strategy. Otherwise, the transmitters do not utilize coordination, all available transmitter sensors are used, and the problem becomes an instance of setting II: a saddle-point solution does not exist and the Stackelberg equilibrium is achieved by deterministic linear strategies. 

Our analysis has uncovered an interesting result regarding coordination among the transmitter sensors and the receiver, and among the adversarial nodes. If the transmitter nodes can coordinate, then the  adversaries will benefit from coordination, i.e., all will generate the identical realization of  an i.i.d. Gaussian noise sequence. If the transmitters cannot coordinate, adversarial sensors do not benefit from coordination, and the resulting Stackelberg equilibrium is at strictly higher cost than the one when transmitters can coordinate (setting I). 

Finally, we have analyzed the impact of optimal power allocation among both the transmitter (defender) and the adversarial (attacker) sensors when various parameters that define the game are not the same for all sensors--the asymmetric case. We have shown that the optimal attack strategy, when the defender can coordinate, allocates all attack power on the best sensor, where the criteria of the selection of best sensor pertains to the receiver SNR.  Moreover, the flexibility of power allocation renders coordination superfluous for the adversarial sensors, while it remains beneficial for the transmitter sensors. In the absence of coordination, both the optimal transmitter and the optimal attacker strategies use all available sensors to distribute power optimally.

Several questions still remain open and are currently under investigation, including  extensions of the analysis to vector sources and channels. The information-theoretic analysis of such a setting  requires a vector form of Witsenhausen's Lemma, which is an important research question in its own right, see \cite{tian2015matched} for recent progress in this direction.  The investigation of optimal power allocation strategies for asymmetric settings for vector sources and channels, and the scaling analysis, in terms of the number sensors, are parts of our current research.  

\begin{appendices}
\section{The Gaussian CEO Problem}
In the Gaussian CEO problem, an underlying Gaussian source $S \sim \mathcal N(0,\sigma_S^2) $ is observed under additive noise $\boldsymbol W \sim \mathcal N(\boldsymbol 0, R_W)$ as $\boldsymbol U=S+\boldsymbol W$. These noisy observations, i.e., $\boldsymbol U$, must be encoded in such a way that the decoder produces a good approximation to the original underlying source. This problem was proposed in \cite{viswanathan1997quadratic} and solved in \cite{oohama2005rate} (see also \cite{oohama1998rate,prabhakaran2004rate}). A lower bound for this function for the non-Gaussian sources within the ``symmetric" setting where all $U$'s have identical statistics was presented in \cite{gastpar2005lower}. Here, we simply extend the results in \cite{oohama1998rate} to our  setting, noting 
\begin{align}
D&=\mathbb E \{(S-\hat S)^2\}, \label{apbir}\\
R&=\min I(\boldsymbol U; \hat S),\label{apiki}
\end{align}
where $\boldsymbol U=\boldsymbol \beta S+\boldsymbol W$,   $\boldsymbol W \sim \mathcal N(\boldsymbol 0, R_W)$, and $R_W$ is an  $M \times M$ identity matrix. The minimization in (\ref{apiki}) is over all conditional densities $p(\hat s|\boldsymbol u)$ that satisfy (\ref{apbir}). The MSE distortion can be written as sum of two terms 
\begin{align}
D=&\mathbb E\{(S-T+ T-\hat S)^2\}=\mathbb E\{(S-T)^2\} +\mathbb E\{(T-\hat S)^2\}\label{apuciki},
\end{align}
where $T\triangleq \mathbb E \{S| \boldsymbol U\}$. Note that  (\ref{apuciki}) holds since 
\begin{equation}\label{apdort}
\mathbb E\{(S-T)(\hat S-T)\}=0,
\end{equation} 
as the estimation error, $S-T$ is orthogonal to any function\footnote{Note that $\hat S$ is also a deterministic function of $\boldsymbol U$, since the optimal reconstruction can always be achieved by deterministic codes.} of the observation, $\boldsymbol U$.  The estimation error $D_{est} \triangleq \mathbb E\{(S-T)^2\} $ is constant with respect to $p(\hat s|\boldsymbol u)$, {\it i.e.}, a fixed function of $\boldsymbol U$ and $S$. Hence, the minimization is over the densities that satisfy a distortion constraint of the form $\mathbb E\{(T-\hat S)^2\} \leq D_{rd}$ and  $R=\min I(\boldsymbol U; \hat S)$. Hence, we write (\ref{apuciki}) as
\begin{equation}\label{apbes}
D=D_{rd}+D_{est}.
\end{equation}
Note that due to their Gaussianity, $T$ is a sufficient statistic of $\boldsymbol U$ for $S$, {\it i.e.}, $S-T-\boldsymbol U$ forms a Markov chain in that order and  $T\sim \mathcal N(0,\sigma_T^2)$. Hence, $R=\min I(\boldsymbol U; \hat S)=\min I(T; \hat S)$ where minimization is over $p({\hat s}|t)$ that satisfy $\mathbb E\{(T-\hat S)^2\} \leq D_{rd}$, where all variables are Gaussian. This is the classical Gaussian rate-distortion problem, and hence:
\begin{equation}\label{apalti}
D_{rd}(R)=\sigma_T^2 2^{-2R}.
\end{equation}
Note that $T=R_{SU} R_U^{-1} \boldsymbol U$, where $R_{SU} \triangleq \mathbb E \{S \boldsymbol U^T\}$ and  $R_U\triangleq \mathbb E \{\boldsymbol U \boldsymbol U^T\}$ which can  be written explicitly as: 
\begin{equation} \label{myru}
R_U= \left( \begin{array}{cccc } 1+\beta_1^2  &\beta_1\beta_2  & \dots &\beta_1 \beta_M  \\ \beta_1\beta_2 & 1+\beta_2^2  & \ldots & \beta_2 \beta_M\\  \vdots & & \ddots & \vdots \\ \beta_1 \beta_M & \ldots& & 1+\beta_M^2\end{array}\right ). 
\end{equation}
Since $R_U$ is structured, it can easily be manipulated. In particular,  $R_U$ admits an eigen-decomposition $R_U=Q_U^T \Lambda Q_U$ where $Q_U$ is unitary  and $\Lambda$ is a diagonal matrix with elements $1, \hdots, 1, 1+\sum_m \beta_m^2$. We compute $\sigma_T^2$  as
\begin{align}
\sigma_T^2&=R_{SU} R_U^{-1} R_{SU}^T= \sigma_S^2\frac{\sum \limits_{m=1}^{M} \beta_m^2 }{1+\sum \limits_{m=1}^{M} \beta_m^2}, \label{ap9}
\end{align}
and using standard linear estimation principles, we obtain 
\begin{equation}\label{ap10}
D_{est}=\sigma_S^2\frac{1 }{1+\sum \limits_{m=1}^{M} \beta_m^2}.
\end{equation}
Plugging (\ref{ap10}) in (\ref{apalti}) and using (\ref{apbes}) yields
\begin{equation}\label{ap11}
D=\sigma_S^2 \left (\frac{1}{1+\sum \limits_{m=1}^{M} \beta_m^2}  +\frac{\sum \limits_{m=1}^{M} \beta_m^2 }{1+\sum \limits_{m=1}^{M} \beta_m^2} 2^{-2R} \right).
\end{equation}


\section{Witsenhausen's Lemma}

In this section, we recall Witsenhausen's lemma \cite{witsenhausen1975sequences}, which is used in the proof of Theorem 2.

\begin{lemma}
Consider a pair of random variables $X$ and $Y$, generated from a joint density $P_{X,Y}$, and two (Borel measurable) arbitrary functions $f,g:\mathbb R\rightarrow \mathbb R$ satisfying 
\begin{IEEEeqnarray}{rrcll}
\mathbb E\{f(X)\}&=&\mathbb E\{g(Y)\}&=&0, \\
\mathbb E\{f^2(X)\}&=&\mathbb E\{g^2(Y)\}&=&1.
\end{IEEEeqnarray}  
Define 
\begin{equation}
\rho^*\triangleq\sup_{f,g} \mathbb E\{ f(X) g(Y)\}
\end{equation} 
Then for any (Borel measurable) functions $f_N, g_N:\mathbb R^N\rightarrow \mathbb R$ satisfying  
\begin{IEEEeqnarray}{rcl}
\mathbb E\{f_N(\boldsymbol X)\}&=&\mathbb E\{g_N(\boldsymbol Y)\}=0 ,\\
\mathbb E\{f_N^2(\boldsymbol X)\}&=&\mathbb E\{g_N^2(\boldsymbol Y)\}=1,
\end{IEEEeqnarray}
for  length $N$ vectors sampled from the independent and identically distributed random sequences $\{X(i)\}$ and $\{Y(i)\}$,where  each $X(i), Y(i)$ pair is generated from $P_{X,Y}$, as $\boldsymbol X=\{X(i)\}_{i=1}^N$ and $\boldsymbol Y=\{Y(i)\}_{i=1}^N$, we have 
\begin{IEEEeqnarray}{rcl}
\sup_{f_N,g_N} \mathbb E\{f_N(\boldsymbol X )g_N(\boldsymbol Y)\} &\,\,\,\leq\,\,\, & \rho^*.
\end{IEEEeqnarray}
Moreover, the supremum and the infimum above are attained by linear mappings, if  $P_{X, Y}$ is a bivariate normal density. 
\end{lemma}
%

\end{appendices}

\bibliographystyle{IEEEbib}

\bibliography{ref}

\begin{thebibliography}{10}

\bibitem{kim2012cyber}
K.~Kim and P.~R. Kumar,
\newblock ``Cyber--physical systems: {A} perspective at the centennial,''
\newblock {\em Proceedings of the IEEE, Special Centennial Issue}, vol. 100,
  pp. 1287--1308, 2012.

\bibitem{sandberg2015cyberphysical}
H.~Sandberg, S.~Amin, and H.~Johansson,
\newblock ``Cyberphysical security in networked control systems: An
  introduction to the issue,''
\newblock {\em IEEE Control Systems}, vol. 35, no. 1, pp. 20--23, 2015.

\bibitem{fawzi2014secure}
H.~Fawzi, P.~Tabuada, and S.~Diggavi,
\newblock ``Secure estimation and control for cyber-physical systems under
  adversarial attacks,''
\newblock {\em IEEE Transactions on Automatic Control}, vol. 59, no. 6, pp.
  1454--1467, 2014.

\bibitem{pasqualetti2013attack}
F.~Pasqualetti, F.~D{\"o}rfler, and F.~Bullo,
\newblock ``Attack detection and identification in cyber-physical systems,''
\newblock {\em IEEE Transactions on Automatic Control}, vol. 58, no. 11, pp.
  2715--2729, 2013.

\bibitem{mo2014detecting}
Y.~Mo, R.~Chabukswar, and B.~Sinopoli,
\newblock ``Detecting integrity attacks on {SCADA} systems,''
\newblock {\em IEEE Transactions on Control Systems Technology}, vol. 22, no.
  4, pp. 1396--1407, 2014.

\bibitem{lamport1982byzantine}
L.~Lamport, R.~Shostak, and M.~Pease,
\newblock ``The {B}yzantine generals problem,''
\newblock {\em ACM Transactions on Programming Languages and Systems (TOPLAS)},
  vol. 4, no. 3, pp. 382--401, 1982.

\bibitem{dolev1982byzantine}
D.~Dolev,
\newblock ``The {B}yzantine generals strike again,''
\newblock {\em Journal of Algorithms}, vol. 3, no. 1, pp. 14--30, 1982.

\bibitem{6582732}
A.~Vempaty, L.~Tong, and P.~K. Varshney,
\newblock ``Distributed inference with {B}yzantine data: {S}tate-of-the-art
  review on data falsification attacks,''
\newblock {\em IEEE Signal Processing Magazine}, vol. 30, no. 5, pp. 65--75,
  Sept 2013.

\bibitem{kosut2008distributed}
O.~Kosut and L.~Tong,
\newblock ``Distributed source coding in the presence of {B}yzantine sensors,''
\newblock {\em IEEE Transactions on Information Theory}, vol. 54, no. 6, pp.
  2550--2565, 2008.

\bibitem{akyol}
E.~Akyol, C.~Langbort, and T.~Ba\c{s}ar,
\newblock ``Information-theoretic approach to strategic communication as a
  hierarchical game,''
\newblock {\em Proceedings of the IEEE}, vol. 105, no. 2, pp. 205--218, Feb
  2017.

\bibitem{langner2011stuxnet}
R.~Langner,
\newblock ``Stuxnet: {D}issecting a cyberwarfare weapon,''
\newblock {\em IEEE Security \& Privacy}, vol. 9, no. 3, pp. 49--51, 2011.

\bibitem{asilomar16}
E.~Akyol, C.~Langbort, and T.~Ba\c{s}ar,
\newblock ``Strategic communication in multi-agent networks,''
\newblock in {\em Proceedings of the IEEE Asilomar Conference on Signals,
  Systems and Computers, 2016}. IEEE.

\bibitem{gastpar2005power}
M.~Gastpar and M.~Vetterli,
\newblock ``Power, spatio-temporal bandwidth, and distortion in large sensor
  networks,''
\newblock {\em IEEE Journal on Selected Areas in Communications}, vol. 23, no.
  4, pp. 745--754, April 2005.

\bibitem{basar1983gaussian}
T.~Ba\c{s}ar,
\newblock ``The {G}aussian test channel with an intelligent jammer,''
\newblock {\em IEEE Transactions on Information Theory}, vol. 29, no. 1, pp.
  152--157, 1983.

\bibitem{basar1985complete}
T.~Ba\c{s}ar and Y.W. Wu,
\newblock ``A complete characterization of minimax and maximin encoder-decoder
  policies for communication channels with incomplete statistical
  description,''
\newblock {\em IEEE Transactions on Information Theory,}, vol. 31, no. 4, pp.
  482--489, 1985.

\bibitem{basar1986solutions}
T.~Ba\c{s}ar and Y.W. Wu,
\newblock ``Solutions to a class of minimax decision problems arising in
  communication systems,''
\newblock {\em Journal of Optimization Theory and Applications}, vol. 51, no.
  3, pp. 375--404, 1986.

\bibitem{bansal1989communication}
R.~Bansal and T.~Ba\c{s}ar,
\newblock ``Communication games with partially soft power constraints,''
\newblock {\em Journal of Optimization Theory and Applications}, vol. 61, no.
  3, pp. 329--346, 1989.

\bibitem{xiao2008linear}
J.~Xiao, S.~Cui, Z.~Luo, and A.~Goldsmith,
\newblock ``Linear coherent decentralized estimation,''
\newblock {\em IEEE Transactions on Signal Processing}, vol. 56, no. 2, pp.
  757--770, 2008.

\bibitem{4915748}
J.~Li and G.~AlRegib,
\newblock ``Distributed estimation in energy-constrained wireless sensor
  networks,''
\newblock {\em IEEE Transactions on Signal Processing}, vol. 57, no. 10, pp.
  3746--3758, Oct 2009.

\bibitem{1597575}
A.~Ribeiro and G.B. Giannakis,
\newblock ``Bandwidth-constrained distributed estimation for wireless sensor
  networks-part i: Gaussian case,''
\newblock {\em IEEE Transactions on Signal Processing}, vol. 54, no. 3, pp.
  1131--1143, March 2006.

\bibitem{1657815}
J.~Jin, A.~Ribeiro, Luo Z.Q., and G.B. Giannakis,
\newblock ``Distributed compression-estimation using wireless sensor
  networks,''
\newblock {\em IEEE Signal Processing Magazine}, vol. 23, no. 4, pp. 27--41,
  July 2006.

\bibitem{4568456}
I.~Bahceci and A.K. Khandani,
\newblock ``Linear estimation of correlated data in wireless sensor networks
  with optimum power allocation and analog modulation,''
\newblock {\em IEEE Transactions on Communications}, vol. 56, no. 7, pp.
  1146--1156, July 2008.

\bibitem{6731588}
F.~Jiang, J.~Chen, and A.L. Swindlehurst,
\newblock ``Optimal power allocation for parameter tracking in a distributed
  amplify-and-forward sensor network,''
\newblock {\em IEEE Transactions on Signal Processing}, vol. 62, no. 9, pp.
  2200--2211, May 2014.

\bibitem{5089496}
H.~Behroozi and M.R. Soleymani,
\newblock ``On the optimal power-distortion tradeoff in asymmetric {G}aussian
  sensor network,''
\newblock {\em IEEE Transactions on Communications}, vol. 57, no. 6, pp.
  1612--1617, June 2009.

\bibitem{gastpar2008uncoded}
M.~Gastpar,
\newblock ``Uncoded transmission is exactly optimal for a simple {G}aussian
  ÒsensorÓ network,''
\newblock {\em IEEE Transactions on Information Theory}, vol. 54, no. 11, pp.
  5247--5251, 2008.

\bibitem{ElGamalBook}
A.~El~Gamal and Y.~Kim,
\newblock {\em Network {I}nformation {T}heory},
\newblock Cambridge {U}niversity {P}ress, 2011.

\bibitem{lapidoth2010sending}
A.~Lapidoth and S.~Tinguely,
\newblock ``Sending a bivariate {G}aussian over a {G}aussian {MAC},''
\newblock {\em IEEE Transactions on Information Theory}, vol. 56, no. 6, pp.
  2714--2752, June 2010.

\bibitem{leong2011scaling}
A.~Leong and S.~Dey,
\newblock ``On scaling laws of diversity schemes in decentralized estimation,''
\newblock {\em IEEE Transactions on Information Theory}, vol. 57, no. 7, pp.
  4740--4759, 2011.

\bibitem{akyol2013gaussian}
E.~Akyol, K.~Rose, and T.~Ba\c{s}ar,
\newblock ``Gaussian sensor networks with adversarial nodes,''
\newblock in {\em Information Theory Proceedings (ISIT), 2013 IEEE
  International Symposium on}. IEEE, 2013, pp. 539--543.

\bibitem{akyol2013communication}
E.~Akyol, K.~Rose, and T.~Ba{\c{s}}ar,
\newblock ``On communication over {G}aussian sensor networks with adversaries:
  Further results,''
\newblock in {\em Decision and Game Theory for Security}, pp. 1--9. Springer,
  2013.

\bibitem{akyol2016power}
E.~Akyol and U.~Mitra,
\newblock ``Power-distortion metrics for path planning over {G}aussian sensor
  networks,''
\newblock {\em IEEE Transactions on Communications}, vol. 64, no. 3, pp.
  1220--1231, 2016.

\bibitem{basarbook}
T.~Ba\c{s}ar and G.~Olsder,
\newblock {\em Dynamic Noncooperative Game Theory},
\newblock Society for Industrial Mathematics (SIAM) Series in Classics in
  Applied Mathematics, 1999.

\bibitem{diggavi2001worst}
S.N. Diggavi and T.M. Cover,
\newblock ``The worst additive noise under a covariance constraint,''
\newblock {\em IEEE Transactions on Information Theory}, vol. 47, no. 7, pp.
  3072--3081, 2001.

\bibitem{boyd2004convex}
S.~Boyd and L.~Vandenberghe,
\newblock {\em Convex Optimization},
\newblock Cambridge University Press, 2004.

\bibitem{tian2015matched}
C.~Tian, J.~Chen, S.~Diggavi, and S.~Shamai,
\newblock ``Matched multiuser {G}aussian source-channel communications via
  uncoded schemes,''
\newblock in {\em 2015 IEEE International Symposium on Information Theory
  (ISIT)}. IEEE, 2015, pp. 476--480.

\bibitem{viswanathan1997quadratic}
H.~Viswanathan and T.~Berger,
\newblock ``The quadratic {G}aussian {CEO} problem,''
\newblock {\em IEEE Transactions on Information Theory}, vol. 43, no. 5, pp.
  1549--1559, 1997.

\bibitem{oohama2005rate}
Y.~Oohama,
\newblock ``Rate-distortion theory for {G}aussian multiterminal source coding
  systems with several side informations at the decoder,''
\newblock {\em IEEE Transactions on Information Theory}, vol. 51, no. 7, pp.
  2577--2593, 2005.

\bibitem{oohama1998rate}
Y.~Oohama,
\newblock ``The rate-distortion function for the quadratic {G}aussian {CEO}
  problem,''
\newblock {\em IEEE Transactions on Information Theory}, vol. 44, no. 3, pp.
  1057--1070, 1998.

\bibitem{prabhakaran2004rate}
V.~Prabhakaran, D.~Tse, and K.~Ramachandran,
\newblock ``Rate region of the quadratic {G}aussian {CEO} problem,''
\newblock in {\em Proceedings of the International Symposium on Information
  Theory}. IEEE, 2004, p. 119.

\bibitem{gastpar2005lower}
M.~Gastpar,
\newblock ``A lower bound to the {AWGN} remote rate-distortion function,''
\newblock in {\em IEEE 13th Workshop on Statistical Signal Processing}. IEEE,
  2005, pp. 1176--1181.

\bibitem{witsenhausen1975sequences}
H.S. Witsenhausen,
\newblock ``On sequences of pairs of dependent random variables,''
\newblock {\em SIAM Journal on Applied Mathematics}, pp. 100--113, 1975.

\end{thebibliography}

\end{document}